\documentclass[aps,prd,preprint,superscriptaddress,nofootinbib]{revtex4}
\pdfoutput=1

\usepackage{color}
\usepackage[pdftex]{graphicx}
 \usepackage{hyperref}

\newcommand{\lp}{\left(}
\newcommand{\rp}{\right)}
\newcommand{\lb}{\left[}
\newcommand{\rb}{\right]}

\newcommand{\ba}{\begin{eqnarray}}
\newcommand{\ea}{\end{eqnarray}}
\newcommand{\be}{\begin{equation}}
\newcommand{\ee}{\end{equation}}

\newcommand{\ga}{\gamma}

\newcommand{\mC}{\mathcal{C}}

\begin{document}

\title{Matters on a moving brane}

\author{Tomi Sebastian Koivisto}\email{tomi.koivisto@fys.uio.no}
\affiliation{Institute of Theoretical Astrophysics, University of Oslo, P.O. Box 1029 Blindern, N-0315 Oslo, Norway}
\author{Danielle Elizabeth Wills}\email{d.e.wills@durham.ac.uk}
\affiliation{Centre for Particle Theory, Department of Mathematical Sciences, Durham University, South Road, Durham, DH1 3LE, UK}

\date{\today}

\begin{abstract}
``An idle brane is the devil's workshop.''
\newline
\newline
A novel generalisation of the Dirac-Born-Infeld string scenario is described. It is shown that matter residing on the moving brane is dark and has the so-called disformal coupling to gravity. This gives rise to cosmologies where dark matter stems from the
oscillations of the open strings along the brane and the transverse oscillations result in dark energy.
Furthermore, due to a new screening mechanism that conceals the fifth force from local experiments, one may even entertain the possibility that the visible sector is also moving along the extra dimensions.
\newline
\newline
\newline
\newline
Essay written for the Gravity Research Foundation 2013 Awards for Essays on Gravitation, Submitted 31.3.2013
\end{abstract}

\maketitle

\newpage

The relation between physical and gravitational geometry is a basic issue of fundamental importance to both classical and quantum theory.
Gravity is the geometry of space-time described by the metric $g_{\mu\nu}$, which in General Relativity is prescribed dynamics by the
Einstein-Hilbert action. The geometry governing the movement of matter fields in the space-time is called the physical geometry $\bar{g}_{\mu\nu}$,
\be \label{action}
S = \int d^4 x \sqrt{-g} \frac{R}{16\pi G}  + S(\mbox{matter},\bar{g}_{\mu\nu})\,.
\ee
Deviations from the postulate of minimal coupling, $\bar{g}_{\mu\nu}=g_{\mu\nu}$, have to be constrained experimentally \cite{will1993theory}.
For simplicity, one may consider such deviations
to be given by a single scalar field $\phi$. It can then be argued that the most general physically meaningful relation between the two metrics has the form \cite{Bekenstein:1992pj}
\be \label{disformal}
\bar{g}_{\mu\nu}=C(\phi,X)g_{\mu\nu}+D(\phi,X)\phi_{,\mu}\phi_{,\nu}\,,
\ee
where $X=(\partial\phi)^2$ is the kinetic term of the field. The function $C$ gives the very well known conformal transformation. In this essay we discuss the role of the function $D$ that gives the so-called disformal transformation.

It is easy to argue that though largely neglected in the past literature, this relation is generic.
The Brans-Dicke class of scalar-tensor theories is described by a nontrivial function $C$; a very extensively studied example is the $f(R)$ type of
gravity theory. However, when one considers {\it any} more general scalar-tensor theory, or writes a covariant action for the metric involving
{\it any} other invariant besides $R$, the function $D$ must appear in the Einstein frame formulation of the theory\footnote{For a unified view of the $f(R)$ gravities and the role of conformal transformations therein, see \cite{Amendola:2010bk}. About the disformal relation in the context of general Horndenski classes of scalar-tensor theories, see \cite{Zumalacarregui:2012us}.}. Here our starting point, instead of an {\it ad hoc} modification of gravity, is
a higher dimensional theory with pure Einstein gravity minimally coupled to matter: we will see that the resulting four-dimensional description generically involves a nontrivial function $D$.

The primary novel contribution of this essay is to present a generalisation of the Dirac-Born-Infeld (DBI) string scenario, that predicts matter couplings
of the form (\ref{disformal}), with both functions $C$ and $D$ robustly derivable from first principles. It is shown that any matter residing upon the moving
brane will have this coupling to our gravitational geometry, when the scalar field $\phi$ is then identified with the radial coordinate of the brane in a warped background, the DBI
radion.
If the brane doesn't intersect the Standard Model stack of branes\footnote{To accommodate the non-Abelian gauge fields of the Standard Model,  one needs a stack of branes instead of just one.}, the disformally coupled matter would interact only gravitationally with baryonic matter.
In particular we propose to associate the $U(1)$ gauge field upon the moving brane with this invisible matter.

\begin{figure}[]
\begin{center}
  \includegraphics[width=0.6\textwidth]{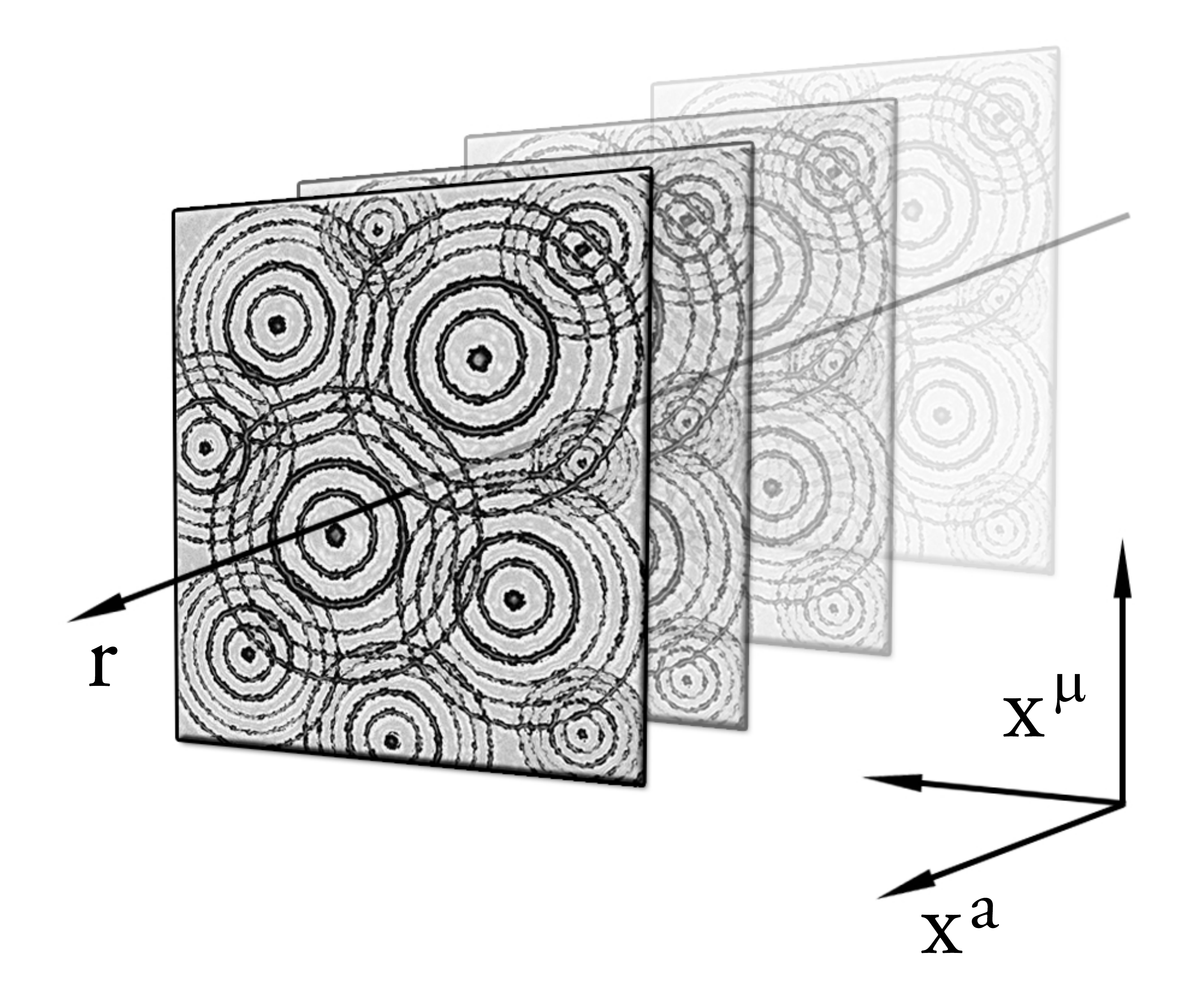}
\end{center}
\caption{Dark matter emerges from oscillations of the open strings along the brane, visualised here as ripples on the surface sourced by the open string endpoints. Dark energy emerges from the oscillations of the open strings transverse to the brane, visualised here as the movement of the brane as a single wave-front or wave-pulse in the radial direction. \label{rbi}}
\end{figure}

An appealing unification of the cosmological dark sector has then emerged: dark matter stems from the oscillations of the open strings in the directions along the brane, which propagate as a vector field on the world-volume, and the dynamical dark energy field is the scalar DBI radion, which encodes the oscillations of open strings transverse to the brane, see Fig. \ref{rbi}. Gravity emerges from the oscillations of the closed strings in the bulk space-time. 
In generic compactifications, one obtains both massive and massless vector fields. In cosmology, the former manifest as dark matter and the latter as ``dark radiation'', both of which are non-minimally coupled to the dark energy field via the disformal relation. 

In a warped flux compactification of Type IIB string theory, the ten-dimensional metric takes the form
\be
G_{MN}dx^Mdx^N = h^{-1/2}g_{\mu\nu}dx^\mu dx^\nu + h^{1/2}g_{ab}dx^a dx^b,
\ee
where $h = h(x^a)$ is the warp factor which depends only on the compact coordinates indexed by $a,b = 4,...,9$.  These are the wrapped dimensions of the Calabi-Yau space. The capital indices $M,N = 0,...,9$ run over all space-time dimensions, the greek $\mu,\nu = 0,...,3$ over the large 4 dimensions.
For simplicity we consider a probe D$3$-brane embedded in this background, as the set-up can readily be generalised to branes of lower codimension. In the Einstein frame the Dirac-Born-Infeld (DBI) action describing the dynamics of a D$3$-brane is given by
\be \label{DBIaction}
S_{\rm DBI}= -(2\pi)^{-3}\ell_s^{-4} \int d^4 \xi \sqrt{-\det(\gamma_{\mu\nu}+e^{-\frac{\varphi}{2}}\mathcal{F}_{\mu\nu})}\,,
\ee
where the integration is over the brane coordinates on the world-volume, $\xi^\mu$, and the string scale $\ell_s^2$ gives the tension of the brane. The dilaton $\varphi$ we assume to be stabilized as usual.
The induced metric on the brane is denoted by $\ga_{\mu\nu}$. Finally, there appears
\be
\mathcal{F}_{\mu\nu}=\mathcal{B}_{\mu\nu}+ 2\pi\ell_s^2 F_{\mu\nu}\,,
\ee
the gauge invariant combination of the pullback of the NSNS 2-form field $\mathcal{B}_2$ and the field strength $F_{\mu\nu}$ of the world-volume $U(1)$ gauge field. Below we will describe how the induced metric $\gamma_{\mu\nu}$ gives rise to the disformal relation in gravity, and the
field $\mathcal{F}_{\mu\nu}$ to disformally coupled matter.

\begin{figure}[h]
\begin{center}
  \includegraphics[width=0.8\textwidth]{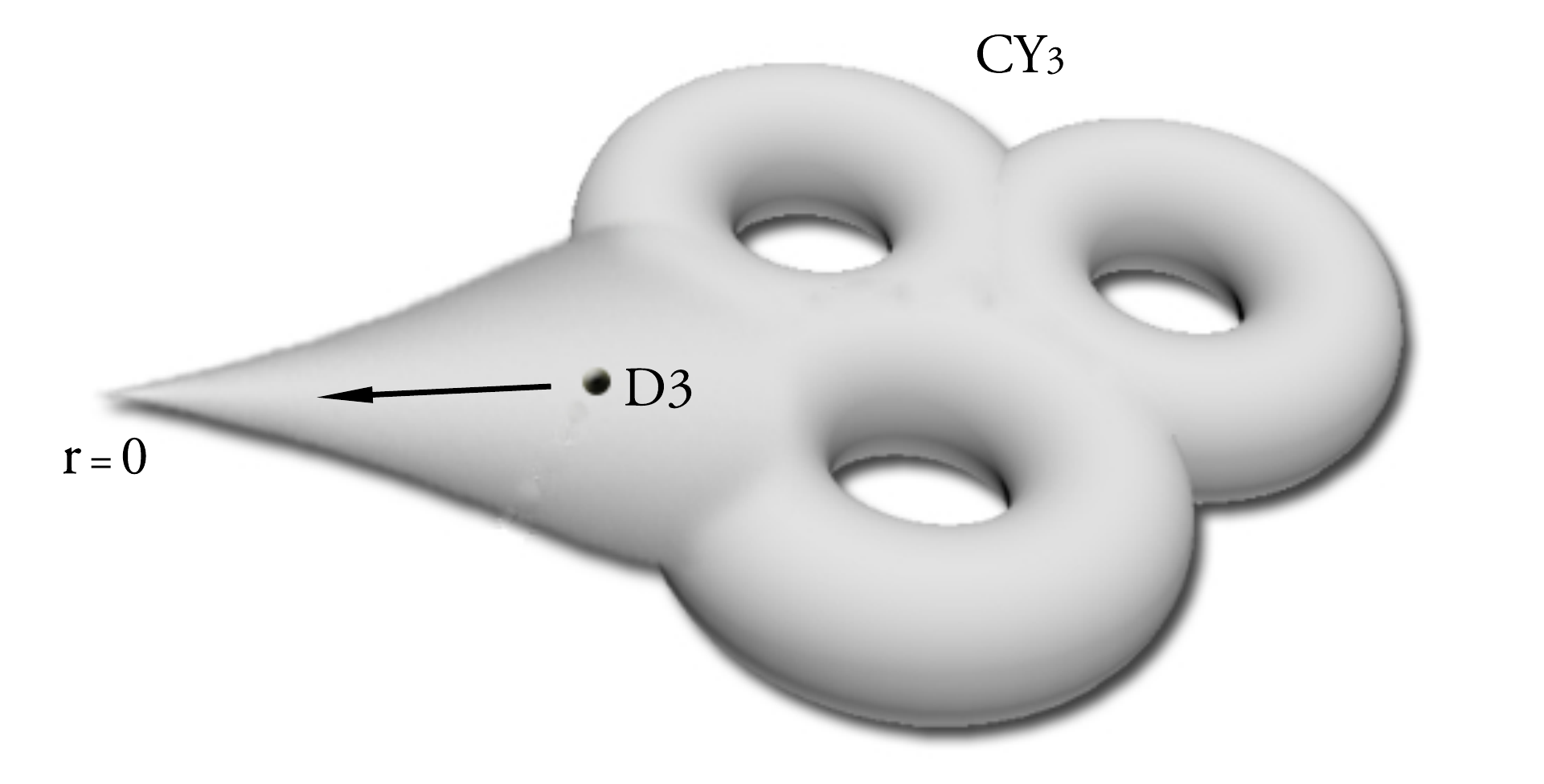}
\end{center}
\caption{A D3-brane moving in the radial direction of a warped throat in a compact Calabi-Yau threefold. \label{rbi2}}
\end{figure}

Let us first focus on the geometry, see Fig. \ref{rbi2}. The pullback of the ten dimensional metric onto the brane world-volume is given by
\be
\gamma_{\mu\nu}= G_{MN}\partial_\mu x^M \partial_\nu x^N\,.
\ee
Adopting the static gauge, the D$3$-brane fills the large dimensions such that $\xi^\mu=x^\mu$.
We allow the compact spacetime coordinates which are transverse to the brane to be functions of the four-dimensional world-volume coordinates, $x^a = x^a(\xi^\mu)$.
This entails that the brane may move in the internal space. With these choices, the four dimensional components of the induced metric on the brane world-volume take the form
\be\label{gamma4D}
\gamma_{\mu\nu} = G_{\mu\nu} + \frac{\partial x^a}{\partial \xi^\mu}\frac{\partial x^b}{\partial \xi^v}G_{ab}\,.
\ee
It can be justified to consider purely one-dimensional brane trajectories in the radial direction of a warped throat \cite{Easson:2007dh}.
Returning to Eq. (\ref{disformal}), we may identify the scalar field $\phi$ with the canonically normalised radial coordinate of the brane, one of the $x^a$'s in Eq. (\ref{gamma4D}). Thus we have obtained the relation (\ref{disformal}), where $\gamma_{\mu\nu} \rightarrow \bar{g}_{\mu\nu}$ and the functions $C \sim h^{\frac{1}{2}}(\phi)$ and $D \sim h^{-\frac{1}{2}}(\phi)$ are given by the warp factor.
\newline

It is straightforward to compute the determinant in the DBI action explicitly to separate the geometric part into the contributions from the gravitational metric and the scalar field. To complete the picture, we include the D3-brane charge $\mC_4$ which is given by the warp factor as $\sim 1/h$, as well as the potential $V(\phi)$ that may emerge from the coupling of the brane to other sectors and is in principle computable given any explicit set-up. We arrive at the following description of our four-dimensional gravity:
\be
S=\int d^4 x \sqrt{-g}\lb \frac{R}{16\pi G} -  \frac{1}{h(\phi)}\lp \sqrt{1+h(\phi)X}-1\rp -V(\phi)\rb  + S\lp\mathcal{F}_{\mu\nu}, \bar{g}_{\mu\nu}\rp\,,
\ee
where all the $\mathcal{F}$-dependent terms are collected into an effective matter action. Here we have written $ \gamma_{\mu\nu}=\bar{g}_{\mu\nu}$, to emphasize that what we have is precisely a nontrivial realization of our starting point (\ref{action}), supplemented with the Lagrangian for the scalar field that governs the relation (\ref{disformal}).

Let us then consider the matter sources. The vector fields are associated with the open string endpoints on the brane,
and may acquire masses via St\"uckelberg couplings to bulk two-forms. The masses will depend on the geometry and are inherently very large, such that for the late-time universe these fields will have long decayed into lighter particles, unless they are stable. In the end, for either case, we do expect massive particles on the brane, and by construction, they will be both dark and disformally coupled to gravity.

Thus we have arrived at a novel generalisation of the coupled quintessence cosmology. In such models, the fine-tuning problem
of dark energy could be alleviated by the fact that the energy density is promoted to a dynamical field, and the coincidence problem could
be addressed by the direct coupling between the dark matter and dark energy components. 
A crucial feature is that the coupling will affect the formation of structure in dark matter and thus will modify the predictions for the observed large-scale structure. 
In the equations governing the evolution of cosmological perturbations, the disformal coupling has been shown to have a considerably richer structure than the purely conformal interaction term \cite{Zumalacarregui:2012us}, and we expect that the precise model presented here can be very efficiently constrained by studying the matter power spectrum at both linear and nonlinear regimes.

An even more radical possibility emerges due to the recently discovered disformal screening mechanism \cite{Koivisto:2012za}, which can conceal the coupling from local experiments even while it has drastic consequences at cosmological scales. In the DBI string scenario presented above this suggests that our Standard Model could be residing on a moving stack of branes\footnote{One may generalise the DBI action to the non-Abelian case.} without introducing a cosmological moduli problem.
This could potentially bring new classes of string cosmologies with novel late-time phenomenology into serious consideration.

The rich phenomenology of disformal couplings has very recently attracted growing interest.
Whereas the function $C$ in (\ref{disformal}) is a local scale transformation that leaves the causal structure untouched, the function $D$ affects angles and thus distorts the light cones. Therefore coupling the electromagnetic field disformally results in a varying-speed-of light theory.   Constraints on such couplings have been derived from both high-precision laboratory experiments of low-energy photons \cite{Brax:2012ie} and from the cosmological evolution of the cosmic microwave background black-body radiation \cite{vandeBruck:2013yxa}. Bounds can also be derived by considering the coupling of baryonic fluids in the radiation dominated epoch of the universe evolution \cite{vandeBruck:2012vq}. As we have shown here, such phenomenology can in fact be considered as a probe of extra dimensional brane movement in a robust string theoretical setting. 

A note added: The theoretical underpinnings and cosmological implications of the scenario outlined here are explored much
further in Ref. \cite{Koivisto:2013fta} where we refer the reader for more details. See also \cite{Bettoni:2013diz,Zumalacarregui:2013pma,Brax:2013nsa} for related recent studies of disformal couplings.

\acknowledgements

TK is supported by the Research Council of Norway and DW by an STFC studentship. We would like to thank Ruth Gregory, David Mota, Ivonne Zavala and Miguel Zumalac\'arregui for useful discussions. 

\bibliography{essayrefs}

\end{document}